\documentclass[aps,prd,reprint,showpacs,twocolumn,letterpaper,superscriptaddress,nofootinbib]{revtex4-1}

\usepackage{graphicx,subfigure}
\usepackage{amsmath,amssymb}
\usepackage{natbib}
\usepackage{bm}
\usepackage{color}
\newcommand{\be}{\begin{equation}}
\newcommand{\ee}{\end{equation}}
\newcommand{\ba}{\begin{eqnarray}}
\newcommand{\ea}{\end{eqnarray}}
\newcommand{\dphi}{d \phi}
\newcommand{\bphi}{\bm{\phi}}
\newcommand{\ssec}[1]{\emph{#1} --- }

\begin{document}
\title{Metropolis Monte Carlo on the Lefschetz thimble: application to a one-plaquette model}
\author{Abhishek Mukherjee}
\email[]{mukherjee@ectstar.eu}
\affiliation{ECT$^\star$, Villa Tambosi, I-38123 Villazzano (Trento), Italy}
\affiliation{LISC, Via Sommarive 18, I-38123 Povo (Trento), Italy}
\author{Marco Cristoforetti}
\email[]{mcristofo@ectstar.eu}
\affiliation{ECT$^\star$, Villa Tambosi, I-38123 Villazzano (Trento), Italy}
\affiliation{LISC, Via Sommarive 18, I-38123 Povo (Trento), Italy}
\author{Luigi Scorzato}
\email[]{scorzato@ectstar.eu}
\affiliation{ECT$^\star$, Villa Tambosi, I-38123 Villazzano (Trento), Italy}
\affiliation{LISC, Via Sommarive 18, I-38123 Povo (Trento), Italy}

\begin{abstract}
We propose a new algorithm based on the Metropolis sampling method to perform Monte Carlo integration 
for path integrals in the recently proposed formulation of quantum field theories on the Lefschetz thimble. 
The algorithm is based on a mapping between the curved manifold defined by the Lefschetz thimble
of the full action and the flat manifold associated with the corresponding quadratic action. We discuss
an explicit method to calculate the residual phase due to the curvature of the Lefschetz thimble.
Finally, we apply this new algorithm to a simple one-plaquette model where our results are in perfect agreement with 
the analytic integration. We also show that for this system the residual phase does not represent a sign problem. 
\end{abstract}

\maketitle

\ssec{Introduction}
In the path integral formulation of quantum field theory (QFT), the expectation value of observables 
is written as ratios of multidimensional functional integrals involving the exponential of 
an (effective) action, $S$. 
When $S$ is real, $e^{-S}$ can be interpreted as a probability distribution and the functional integral can
be evaluated very efficiently and accurately using stochastic methods, viz. Monte Carlo sampling (see, e.g., \cite{Morningstar:2007}). 
For large systems at low temperatures, quantum Monte Carlo is arguably the most accurate method for calculating observables, at present.

Unfortunately, systems with real actions are special cases. In general, $S$ will be complex (although the full integral is still
real), and $e^{-S}$ cannot be interpreted as a probability distribution. In principle, one can use reweighting: the absolute value of $e^{-S}$,
i.e., $e^{-\Re{S}}$ as the probability weight and include $e^{-\Im{S}}$ in the redefinition of the value of the observable for a 
given field configuration. However, reweighting is effective only if the fraction of configurations with negative weight is limited,
 rendering the method of little use for large systems and/or at low temperatures.
This is a manifestation of the infamous `sign problem' which plagues the application of Monte Carlo methods to quantum field theories.
 
Numerous methods have been proposed to deal with sign problem \cite{Baer:1998,Aarts:2013bla,Chandrasekharan:2010}, 
and they have had important but  partial success in particular classes of models. 
However, a general solution is missing, and the sign problem is a major hindrance to accurate calculation in many interesting physical systems:
lattice QCD at finite density \cite{Aarts:2013bla} or with a $\theta$-vacuum \cite{Vicari:2008jw}, real-time field theories \cite{Berges:2006xc},
electronic systems \cite{PhysRevLett.71.1148,charutz:4495,Baer:1998} the repulsive Hubbard model \cite{Loh:1990zz},
 the nuclear shell model \cite{Koonin:1996xj}, polymer field theory \cite{Vilgis2000167}, to name a few. Any 
new method to evade or at least mollify  the sign problem in the generic situation represents an important advance.

Recently we proposed that a way to alleviate the sign problem is to use the formulation of the QFT on a Lefschetz thimble \cite{Witten:2010cx,Witten:2010zr}
for the Monte Carlo integration \cite{PhysRevD.86.074506,Cristoforetti:2013wha}. 
Lefschetz thimbles are many dimensional generalizations of the paths of steepest descent.
By construction the imaginary part of the action remains constant on each thimble. However, because the Lefschetz thimbles are
in general curved complex manifolds, we may pick up an additional \emph{residual phase} due this curvature. We argued that 
the sign problem due to this residual phase, if present at all, should be much milder than the sign problem in the original
integration domain.

The Lefschetz thimble formulation of QFT is, in principle, independent from methods used to sample field configurations on the
thimble. The latter, in itself, presents a non-trivial problem due to complexity of the measure on the thimble. In previous work,
we proposed an algorithm based on discretized Langevin dynamics.
While the testing of the algorithm proposed in \cite{PhysRevD.86.074506} is in progress, it is also worth exploring altrnative algorithms to achieve the challenging goal of performing Monte Carlo simulations on a Lefschetz thimble.

In the this paper, we present a different method to sample field configuration on the Lefschetz thimble, which is based
on the Metropolis algorithm and uses a mapping between the Lefschetz thimble and a flat manifold associated with the corresponding
quadratic action. We also discuss an explicit procedure to calculate the residual phase within this method. 

We apply this method to the $U(1)$ one-plaquette model. The integrals involved in this model are one variable integrals and can 
be performed analytically. However, it provides an interesting benchmark which can be seen as a limiting case of more realistic 
QFTs on a lattice. It is non-trivial from the point of view of a Monte Carlo integration. In fact, the complex Langevin method 
fails for this particular system. It also provides a case where different aspects of our methodology can be visualized quite clearly.

\ssec{QFT on a Letschetz  thimble}
Consider a QFT on a lattice (or any other system with a finite number of continuous degrees of freedom) defined by
 the action $S(\bphi)$, where $\bphi$ is a vector field 
whose number of components, $n$, is equal to the number of degrees of freedom in the system. Suppose that the initial field theory is defined for real fields, i.e,
the expectation value of any observable $\mathcal{O}$ is given by,
\be
\langle \mathcal{O}\rangle = \frac{\int_{\mathcal{D}} d\bphi \mathcal{O} (\bphi) e^{- S (\bphi)}}{\int_{\mathcal{D}} d\bphi e^{- S (\bphi)}}
\ee
where $\mathcal{D}$ is the appropriate integration cycle for $S$ in the real domain $\mathbb{R}^n$.
Now, consider $S$ in terms of the complexified fields, i.e, the field components $\phi_i$ are now allowed to be complex. 
Suppose, $S(\bphi)$ is holomorphic in this complexified space
and its critical points $\bphi^{\sigma}$ given by 
\be
\label{eqncr}
\frac{\partial S}{\partial \bphi^{\sigma}} = 0
\ee
are non-degenerate,
\be
\det \left [ \frac{\partial^2 S}{ \partial \bphi^{\sigma} \partial \bphi^{\sigma}} \right ] \neq 0.
\ee 
Then, under suitable conditions on $S$ and $\mathcal{O}$ (typically fulfilled in physical systems) and for a sufficiently generic choice of parameters,
 we have the following crucial result \cite{Witten:2010cx,Witten:2010zr,Pham:1983}
\be
\int_{\mathcal{D}} d\bphi \mathcal{O} (\bphi)  e^{-S(\bphi)} = \sum_{\sigma}m_\sigma\int_{\mathcal{J}_\sigma} d\bphi  \mathcal{O} (\bphi) e^{-S(\bphi)}\;,
\ee
where $m_\sigma\in\mathbb{Z}$ (see later).
That is, an integral over the real domain $\mathcal{D}$ is equivalent to sum of integrals over the Lefschetz thimbles $\mathcal{J}_\sigma$.
This result can be seen as a generalization of contour deformation in one dimension. 
The Lefschetz thimbles $\mathcal{J}_{\sigma}$ associated with the critical points are 
many dimensional generalizations of the paths of steepest descent. The thimble $\mathcal{J}_{\sigma}$ is defined as the union of all paths governed by,
\be
\label{eqnsd}
\frac{d\bphi}{d\tau} = - \overline{\frac{\partial S}{\partial \bphi}}
\ee
and which end at the critical point $\bphi^{\sigma}$ for $\tau \to \infty$. They are hypersurfaces of \emph{real} dimension $n$ embedded 
in the complex manifold $\mathbb{C}^n$. Here, and later, the overhead bar represents complex conjugation
In this paper we will assume that $S$ is a Morse function, i.e., it has only non degenerate critical points\footnote{Degenerate minima, as they typically occur in the presence of symmetries, can be either lifted or treated as discussed in \cite{PhysRevD.86.074506}}.


Then, the expectation value of an observable can be written as
\be
\label{eq:obs_thi}
\langle \mathcal{O} (\bphi) \rangle = \frac{\sum_{\sigma} m_{\sigma}\int_{\mathcal{J_{\sigma}}} d\bphi \mathcal{O} (\bphi) e^{- S (\bphi)}}{\sum_{\sigma} m_{\sigma}\int_{\mathcal{J_{\sigma}}}  d\bphi (\bphi) e^{- S (\bphi)}} \;.
\ee
From the point of view of stochastic integration, the main benefit of the above formulation is that along a given thimble $\mathcal{J}_{\sigma}$, the imaginary part of the action $\Im{S(\bphi)}$ remains constant. The only fluctuation in the complex phase comes from the residual phase due to the curvature of the thimble itself. We expect this to be a significantly milder sign problem than the original one.


The critical points of the action can be found by looking at all the solutions of Eq.~(\ref{eqncr}).
The integer coefficients $m_{\sigma}$ are the intersection numbers between $\mathcal{D}$ and $\mathcal{K}_{\sigma}$, 
where $\mathcal{K}_{\sigma}$ is the unstable thimble,
i.e, it is the union of all paths which are governed by Eq.~(\ref{eqnsd}), but go to $\bphi^{\sigma}$ at $\tau \to -\infty$. It is also 
a hypersurface of real dimension $n$. Then, $m_{\sigma}$ is simply the number of times the two hypersurfaces $\mathcal{D}$ and $\mathcal{K}_{\sigma}$ 
intersect. 



We are not aware of a general method to calculate the $m_\sigma$ for an arbitrary QFT. But we argued in [14] that only a limited set 
of thimbles are expected to dominate and, moreover, a single thimble is typically sufficient to regularize a QFT 
\footnote{See also \cite{Guralnik:2010} for a different point of view, that is complementary and consistent with the one of [14].}. 
 However, in order to test the algorithm presented in this paper, it may be interesting to consider also the case in which we want 
to study more thimbles at the same time. Hence, in the rest of this paper we will keep a general $m_\sigma$, but we will assume 
that the intersection numbers $m_\sigma$ are known, and comment when relevant.

%

\ssec{Mapping the Lefschetz thimble on a flat manifold}
In the neighborhood of a non-degenerate critical point $\bphi^{\sigma}$, the holomorphic action function $S(\bphi)$ can be written  as,
\be
S(\bphi) = S(\bphi^{\sigma})+S_G (\bm{\eta}) + O(|\eta|^3)
\ee
where the Gaussian action $S_G$ is given by,
\be
S_G = \frac{1}{2} \sum_{k} \lambda_k \eta_k^2.
\ee
and $\bm{\eta}$ is related to $\bphi$ by a (complex) linear transformation,
\be
\phi_i = \phi^{\sigma}_i + \sum_k \mathbf{w}_{ki} \eta_k 
\ee
The $\mathbf{w}_{ki}$ are components of the vectors $\mathbf{w}_k$. We call the \emph{flat} thimble associated with the Gaussian action $S_G$, 
the Gaussian thimble $\mathcal{G}_{\sigma}$.

The  $\lambda_k$ and $\mathbf{w}_k$ can be found from the solutions of the generalized eigenvalue equation,
\be
\label{geneig}
\mathbf {H w}_k = \lambda_k \bar{\mathbf{w}}_k \;. 
\ee
The elements of the hessian matrix $\mathbf{H}$ are given by,
\be
H_{ij} = \frac{\partial S}{\partial \phi_i \partial \phi_j }.
\ee
In practice, we find the ${\lambda_k}$ and the $\mathbf{w}_k$ from the positive eigenvalues and the corresponding eigenvectors 
of the real symmetric $2n \times 2n$ matrix 
\be
\mathbf{\tilde{H}} = \left ( \begin{array}{cc} 
                              H^R  & H^I \\
			      H^I  & -H^R

                             \end{array}\right )
\ee
where 
\begin{eqnarray}
  H^R_{ij} &=& \frac{\partial \Re{S}}{\partial{\Re{\phi_i}} \partial {\Re{\phi_j}}} \\
  H^I_{ij} &=& -\frac{\partial \Re{S}}{\partial{\Im{\phi_i}} \partial {\Re{\phi_j}}}. 
\end{eqnarray}
The eigenvalues of $\mathbf{\tilde{H}}$ come in pairs $\{\pm \lambda_k\}$ with $k=1,\ldots n$, and the $\lambda_k$ being real and positive.
Let  $(\mathbf{u}_k$ and $\mathbf{v}_k$ be normalized $n$-dimensional vectors such that $(\mathbf{u}_k^\intercal,\mathbf{v}_k^\intercal)^\intercal$ 
is an eigenvector of $\mathbf{\tilde{H}}$  with a positive eigenvalue $\lambda_k$. Then, the pair $\lambda_k$ and $\mathbf{w}_k = \frac{1}{2}\left ( \mathbf{u}_k+i\mathbf{v}_k \right )$ satisfies Eq.~(\ref{geneig}).

 With this parametrization, the directions of steepest descent/ascent of $\Re{S}$ 
(and constant $\Im{S})$ correspond to directions where the $\eta_k$ are real.
Consider, the equations of steepest descent of the variables $\eta_k$ (assumed real) for the Gaussian 
action $S_G$ in terms of the new parameter $r=e^{-\tau}$,
\be
\frac{d \eta_k} {d r}= \frac{1}{r} \overline {\frac{\partial S_G}{\partial \eta_k}} = \frac{1}{r}\lambda_k \eta_k
\ee
which yields the solution,
\be
\eta_k \propto  r^{\lambda_k}.
\ee

Now, we can define a mapping between the Gaussian thimble, parametrized by the vectors $\bm{\eta}$,  and the Lefschetz thimble, parametrized by the field $\bphi$.
 First, we find the corresponding configuration $\bm{\xi}$ at $r=\epsilon$,
\be
\xi_k =  \epsilon^{\lambda_k} \eta_k.
\ee

For a sufficiently small $\epsilon$, the Lefschetz thimble and the Gaussian thimble will coincide at $r=\epsilon$. 
Thus, the field configuration on the Lefschetz thimble at $r=\epsilon$ is given by,
\be
\begin{aligned}
\phi_i(r=\epsilon) &=\phi^\sigma_i + \sum_k \mathbf{w}_{ki} \xi_k \\
              &=\phi^\sigma_i + \sum_k \epsilon^{\lambda_k} \mathbf{w}_{ki} \eta_k 
\end{aligned}
\ee
Using this as the boundary condition, we can now integrate the equation of steepest descent of the full action
$S$ for the fields $\phi_i (r)$,
\be
\frac{\dphi_i}{dr} = \frac{1}{r} \overline {\frac{\partial S}{\partial \phi_i}}
\ee
from $r=\epsilon$ to $1$. The field configuration at $r=1$ is the one we seek. For brevity, we will simply denote it by $\bphi$.

For a constant $\epsilon$, we have the following relation between the measures of integration 
\be
\label{eqn:mea}
\int_{\mathcal{J}_\sigma}d\bphi = \int_{\mathbb{R}^n} \det \left [ \mathbf{J^{\bphi}_{\bm{\eta}}} \right ] d\bm{\eta} = \int_{\mathbb{R}^n}\displaystyle \left ( \prod_k \epsilon^{\lambda_k} \right )  \det \left [ \mathbf{J^{\bphi}_{\bm{\xi}}} \right ] d\bm{\eta} \;.
\ee
The matrix $\mathbf{J}^{\phi}_{\eta}$ ( $\mathbf{J}^{\phi}_{\xi}$) is the Jacobian of the transformation between the 
$\bm{\eta}$ ($\bm{\xi}$) and $\bphi$ fields.

The matrix $\mathbf{J}^{\phi}_{\xi}$ can be calculated along the path of steepest descent from the equation
\be
\label{eqn:jac1}
\frac{d \left [ \mathbf{J^{\bphi}_{\bm{\xi}}} \right ]_{ik}} {dr} = \frac{1}{r} \overline{\frac{\partial^2 S }{\partial \phi_i \partial \phi_j}} \overline{\left [ \mathbf{J^{\bphi}_{\bm{\xi}}} \right ]}_{jk}
\ee
along with the boundary condition,
\be
\label{eqn:jac2}
\left [ \mathbf{J^{\bphi}_{\bm{\xi}}} \right ]_{ik} (r=\epsilon) = \mathbf{w}_{ki} \;.
\ee

In the limit $\epsilon \to 0$, the above procedure produces an explicit mapping between the flat Gaussian thimble 
and the Lefschetz thimble.
In practice, it necessary to perform calculations at a few sufficiently small values of $\epsilon$ in order to 
perform the extrapolation to the limiting case.
 For later reference, we note that setting $\epsilon = 1$, corresponds
to a mapping from the Gaussian thimble to itself.
 
Note that Eq.~(\ref{eqn:jac1}) involves the evolution of a $N\times N$ matrix whose determinant must also be computed. 
The latter is expected to cost $O(N^3)$.  This may be still too expensive for some models, but it is already a huge 
cost reduction compared to the $O(e^N)$ scaling expected in general and it should be sufficient to enable the Monte 
Carlo simulation of some important models, which are currently not feasible. Techniques of noise estimation of the trace
 (see, e.g., \cite{Dong:1994, Cristoforetti:2013xqcd} )
 may further reduce the cost of the computation of the determinant, but we do not consider them in this paper.

\ssec{Metropolis sampling on the Lefschetz thimble}
Given the mapping above, it is straightforward to formulate a Metropolis algorithm on the Lefschetz thimble.
Below we give the simplest version.

Suppose we start from a set $\{\sigma^{\rm old},\bm{\eta}^{\rm old},\bphi^{\rm old}\}$. First, we propose 
a thimble $\sigma'^{\rm new}$ from the distribution $m_{\sigma'^{\rm new}}/\sum m_\sigma$.
Note that, in view of the arguments presented earlier, this step is typically not needed in simulations of QFT. 
It is done here to compare with the exact analytical result, which is available.
 
Next, we choose $n$ independent standard normal deviates $\{\tilde{\eta}_k\}$.
The $\bm{\eta}'^{\rm new}$ is then obtained as,
\be
\label{eq:etaup}
\eta'^{\rm new}_k = \frac{1}{\sqrt{\lambda_k}} \tilde{\eta}_k.
\ee
Subsequently, $\bphi'^{\rm new}$ is obtained from  $\bm{\eta}'^{\rm new}$ using the procedure outlined above.

The new field configuration is accepted according to the probability,
\be
\label{eq:accp}
P_{\rm accept} = \min \{1, e^{-  \Re S( \bphi'^{\rm new} ) + \Re S( \bphi^{\rm old}) +  S_G(\bm{\eta}'^{\rm new}) -S_G (\bm{\eta}^{\rm old})} \}
\ee


Note that each new configuration proposed in this way is completely independent from the previous ones. The acceptance of such proposals may 
be good as long as the quadratic approximation of the action (that constitutes the basis for the proposal) approximates well the full action. 
 This may not be hopeless, thanks to the basic property of the Lefschetz thimble. In fact, along the thimble, the dominant part of the integral 
is optimally concentrated close to the stationary point. Indeed, this fact was exploited also in \cite{Baer:1998}. In any case, the present 
approach does not rely essentially on the proposal in Eq.~(\ref{eq:etaup}): it is conceivable to devise a proposal based on a Markov chain, by introducing 
small random variations to a previous configuration. The key idea of the present algorithm is rather the mapping between the Lefschetz thimble 
and the Gaussian thimble $\mathcal{G}_\sigma$.

In either case, given a set of $N$ un(de)correlated field configurations labeled by $\alpha=1,\ldots N$, the expectation values of observables are given by 
\be
\langle \mathcal{O} \rangle = \frac {\sum_{\alpha} \mathcal{O}_{\alpha}  J_\alpha e^{-\Im S_{\alpha}}}{\sum_{\alpha} J_\alpha e^{-\Im S_{\alpha}}} 
\ee
where $S_\alpha$, $O_\alpha$ and $J_{\alpha}$ are, respectively, the values of the action, the observable, the determinant of the Jacobian 
defined in Eqs.~(\ref{eqn:mea} - \ref{eqn:jac2}) for the  $\alpha^{\rm th}$ field configuration. Note that, 
although the $\Im S$ remain constant over each thimble, it can vary from thimble to thimble.

This algorithm is inherently stable. As $\epsilon \to 0$, the field configurations will be sampled with the correct measure
on the Lefschetz thimble. At finite $\epsilon$, the distance of sampled field configurations from the Lefschetz 
thimble is \emph{not} accumulated over simulation time and there is no chance of divergences. This is because successive $\bphi$s are
calculated by first generating the $\bm{\eta}$s.  
  
\ssec{One-plaquette model with $U(1)$ symmetry}
We now discuss the application of the above algoritm for a system 
with one degree of freedom, viz. the one-plaquette model with $U(1)$
symmetry.
The action is given in terms of the gauge link $U=e^{i\phi}$ as
\be
S=-i\frac{\beta}{2}\left(U+U^{-1}\right)=-i\beta\cos\phi .
\ee
where $\phi$ in this case is a one component field.
For real $\beta$ the action is complex, similar to real time gauge theories.

For this simple model, all the integrals can be evaluated analytically, which offers the chance to compare every detail of 
our numerical results to exact results. In particular the plaquette average of the phase $e^{i\phi}$ is given by,
\be
\langle e^{i\phi} \rangle = i\frac{J_1(\beta)}{J_0(\beta)} 
\ee
with $J_n(\beta)$ being Bessel functions of the first kind.
This analytic result offers the chance of a clear test of our algorithm.

Obtaining this result using stochastic methods is quite non-trivial.  
For example the complex Langevin method without \emph{ad-hoc} optimizations gives the wrong result for this model \cite{Berges:2007nr}.

In order to apply our method, we treat the field $\phi$ as complex. The action $S$ has two critical points at $\phi=0$ and $\pi$.
By explicitly constructing the Hessian, it is easy to show that both the critical points are non-degenerate.
In this simple model we can also compute the intersection numbers ($m_{\sigma}$), which turn out to be equal to 1 for both thimbles.
The field configurations on the two thimbles are related by the discrete symmetry transformation $\phi \to \pi - \overline{\phi}$,
and expectation values of observables can be written in terms of integrals over one thimble only. However, in order to illustrate the above algorithm,
 we perform stochastic integration using the full Eq.~(\ref{eq:obs_thi}).

\begin{figure}
\includegraphics[width=\columnwidth]{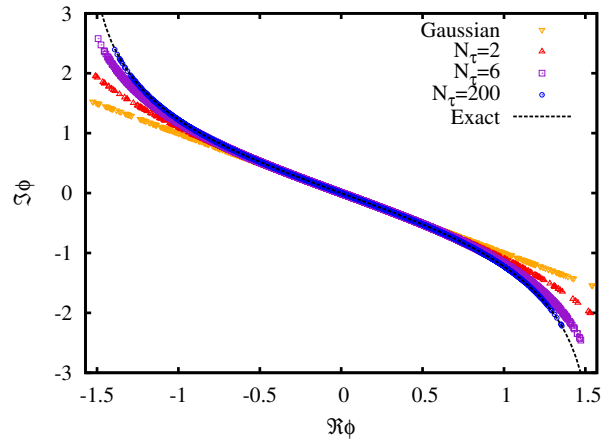}
\caption{Sampled field configurations at $\beta=1$ for the thimble attached to $\phi=0$.}
\label{fig:thimble_conv}
\end{figure}

For this model, one can explicitly derive the expression for the thimbles attached to the two saddle points.
This can be obtained by requiring that the imaginary part of the action be constant along the flow, which gives 
\be
\label{eq:thi}
\cos\Re\phi\cosh\Im\phi= \pm 1
\ee
as the equations for the Lefschetz thimbles attached to the two saddle points.
Such a simple characterization of the thimble is not available for systems with
 more than one degree of freedom. Of course, our algorithm does not make use of Eq.~(\ref{eq:thi}), 
but in Fig.~\ref{fig:thimble_conv} we show that the fields obtained using the method
 described above reproduce well the exact thimble 
defined by Eq.~(\ref{eq:thi}).

We see systematic improvement in our results on increasing $N_{\tau} = \epsilon^{-1}$; with increasing $N_{\tau}$ the sampled 
field configurations uniformly converge on the true thimble. 
In contrast, the flat Gaussian thimble ($N_{\tau} =1$) approximates the thimble quite well near the saddle 
point, but it noticibly different further away from the saddle point.

\begin{figure}
\includegraphics[width=\columnwidth]{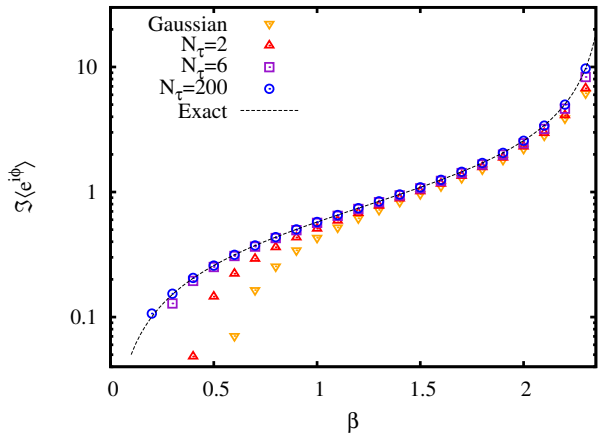}
\caption{Expectation value of $e^{i\phi}$ as a function of $\beta$.}
\label{fig:results_conv}
\end{figure}

In Fig.~\ref{fig:results_conv} we show the results for the expectation value of the observable $e^{i\phi}$ for different $\beta$. 
Again, the results from our method systematically approach the exact analytical result with increasing $N_{\tau}$.
For $N_{\tau}=200$, the results from our method are identical 
(within statistical errors) to the analytical results for the range of $\beta$ considered. In contrast, we notice that 
there is a large difference between the analytical result and those from Monte Carlo if the field configurations are
sampled from the flat Gaussian thimble.

Finally, we discuss the residual phase in the context of the $U(1)$ one-plaquette model.
The question of the residual phase is an important one. 
We expect it to produce a milder sign problem (if at all),
than the original sign problem. Nevertheless, it should be included in any quantitative estimate. 
\begin{figure}
\includegraphics[width=\columnwidth]{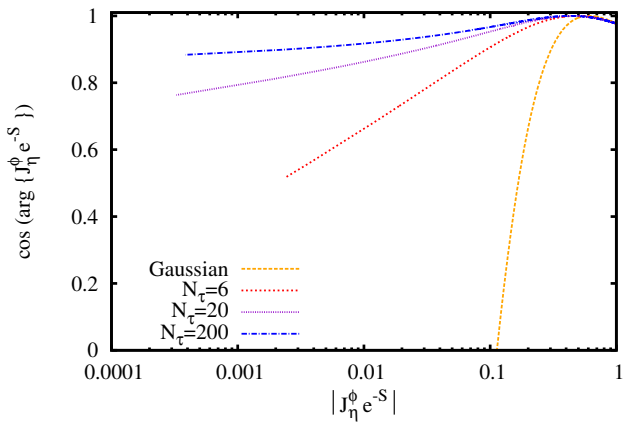}
\caption{The residual phase as a function of the probability measure at $\beta=1$.}
\label{fig:resphase}
\end{figure}
In our formulation the full (complex) measure of integration is given by $\det \left [ \mathbf{J^{\bphi}_{\bm{\eta}}} \right ] e^{-S} $. 
The full integrals on the Lefschetz thimble are always real. This means that
$\sin \left (\arg \left \{\det \left [ \mathbf{J^{\bphi}_{\bm{\eta}}} \right ]  e^{-S} \right \} \right ) $ does not contribute to the integral.
The statement that the sign problem in our method is mild (or absent) means that $\cos \left (\arg \left \{ \det \left [ \mathbf{J^{\bphi}_{\bm{\eta}}} \right ] e^{-S} \right \}\right ) $ (residual phase)
will vary very little (or not at all), in the region where $\left | \det \left [\mathbf{J^{\bphi}_{\bm{\eta}}} \right ] e^{-S} \right | $ (probability measure) is significant.

For the $U(1)$ one-plaquette model, the Jacobian of the transformation on each thimble is a single number and is simply given by,
\be
J^{\phi}_{\eta} =  \frac{-i\beta \, \overline{\sin{\phi}}}{\eta}.
\ee
In Fig.~\ref{fig:resphase} we show the residual phase vs the positive probability measure for this model.
 We see that the residual phase changes by very little 
 for variations of the probability measure spanning many orders of magnitude. Moreover, the fluctuations
of the residual phase grow milder as the true thimble is approached starting from the Gaussian thimble.
Most importantly, the residual phase keeps the same sign throughout the full domain of integration, i.e., 
there is \emph{no sign problem} for our method for this particular model. This is reassuring, although it
 is impossible to extrapolate from this simple model any claim about the residual phase on systems with many degrees of freedom. 

\ssec{Conclusions}
In this paper we have described a new stable algorithm to sample field configurations on the Lefschetz thimble. 
We applied this method to the one plaquette model with $U(1)$ symmetry. Our results are in perfect agreement with
the exact results from analytical integration. Also, the residual phase remains quasi-constant over 
configurations with large weight, indicating that our method does not suffer from a sign problem for this system.
 Further optimization of the algorithm in order to apply it to more challenging
problems with a large number of degrees of freedom is underway. 

\ssec{Acknowledgments}
We would like to thank Francesco Di Renzo, Giovanni Eruzzi, Christian Torrero and Christian Schmidt for useful discussions.

\bibliography{thimble}

\end{document}